\documentclass[a4paper]{jpconf}
\usepackage{graphicx}
\pdfoutput=1
\def\lsim{\mathrel{\rlap{\lower4pt\hbox{\hskip1pt$\sim$}}
    \raise1pt\hbox{$<$}}}         
\def\gsim{\mathrel{\rlap{\lower4pt\hbox{\hskip1pt$\sim$}}
    \raise1pt\hbox{$>$}}}         

\begin{document}
\title{CN Neutrinos and the Sun's Primordial Core Metalicity}

\author{Wick Haxton}

\address{Institute for Nuclear Theory and Department of Physics \\
Box 351550, University of Washington, Seattle WA 98195}

\ead{haxton@phys.washington.edu}

\begin{abstract}
I discuss the use of neutrinos from the CN cycle and pp chain to
constrain the primordial solar  core abundances  of C  and N  at an
interesting level of  precision.  A comparison
of  the  Sun's  deep interior and surface compositions would test  a  key
assumption of the standard solar  model (SSM), a homogeneous zero-age Sun.  It
would   also  provide   a   cross-check  on   recent  photospheric   abundance
determinations that have altered the  once excellent agreement between the SSM
and  helioseismology.   Motivated by the discrepancy between convective-zone
abundances and helioseismology, I discuss the possibility that a two-zone Sun could
emerge from late-stage metal differentiation in the solar nebula connected with 
formation of the gaseous giant planets. 
\end{abstract}

\section{Introduction}
One of the initial motivations  for pursuing solar neutrino physics was
to test our understanding of main-sequence stellar evolution and  the Standard Solar 
Model (SSM).   In the past two decades
this goal was put aside, as difficulties in understanding the pattern of solar neutrino
fluxes led to the discovery of solar neutrino oscillations.  But because of the precision
with which the relevant flavor physics is now known -- and because the solar neutrino
problem also spurred progress in the nuclear physics of the Sun and
the development of high-statistics detectors such as Super-Kamiokande (SK) and
SNO -- the use of neutrinos as quantitative solar probe is now 
a practical possibility.  Specifically, this talk summarizes recent arguments by Aldo Serenelli
and me \cite{whas} that a measurement of the CN neutrino flux
could test a key assumption of the SSM,
the homogeneity of the zero-age main sequence Sun.  This  assumption,
the  basis for equating the SSM's  primordial core metal abundances to  today's surface metal
abundances, may now  be in some
degree of  conflict with observation:  recent 3D  modeling of
photospheric absorption  lines has  led to a  downward revision in  the metal
content  of  the solar  convective  zone  \cite{ags05}, altering
helioseismology  in  the upper radiative zone, where the temperature $\sim$ 2-5 $\times$ 10$^6$
K \cite{bbps05,bsb05,antia,montalban}. In this 
region C, N, O, Ne, and Ar are partially 
ionized  and  particularly  O and  Ne  have  a  significant influence  on  the
radiative opacity. 

A  quantitative comparison between  the Sun's  surface and  core abundances
could prove  useful in understanding  the chemical evolution of  other gaseous
bodies in  our solar system,  whose interiors are  not as readily  probed.  The
Galileo and Cassini  missions found significant metal enrichments  in the H/He
atmospheres of Jupiter and Saturn, e.g.,  abundances of C and N of $\sim$ four
times solar for Jupiter and $\sim$ 4-8 for Saturn \cite{guillot}.  Planetary 
models that  take account of these  data show that the  gaseous giants are
very significant  solar-system metal reservoirs.   The metal excess in these 
planets, estimated to be $\sim~40-90M_\oplus$, is very similar to the depletion that would be needed in the convection zone
to bring the photospheric abundances and helioseismology into agreement.  
Serenelli and I raised the possibility that a two-zone Sun resulted from a process in
which the last $\sim$ 1\% of nebular
gas was scoured of its metals by the process of planetary formation, then deposited
on the early Sun to dilute the convective zone.

\section{The CNO Bi-Cycle and its Neutrinos}
The need for two mechanisms to  burn hydrogen was recognized in the pioneering
work  of  Bethe  and  collaborators.   The first, the pp-chain,  dominates  energy
production  in  our  Sun  and  other  low-mass  main-sequence  stars and can  be
considered a primary process in  which the chain's ``catalysts," deuterium,
$^3$He, and  $^7$Be/$^7$Li (the  elements participating in  the steps
leading to $^4$He), are  synthesized  as  the chain  burns  to
equilibrium.

\begin{figure}
\begin{center}
\includegraphics[width=16cm]{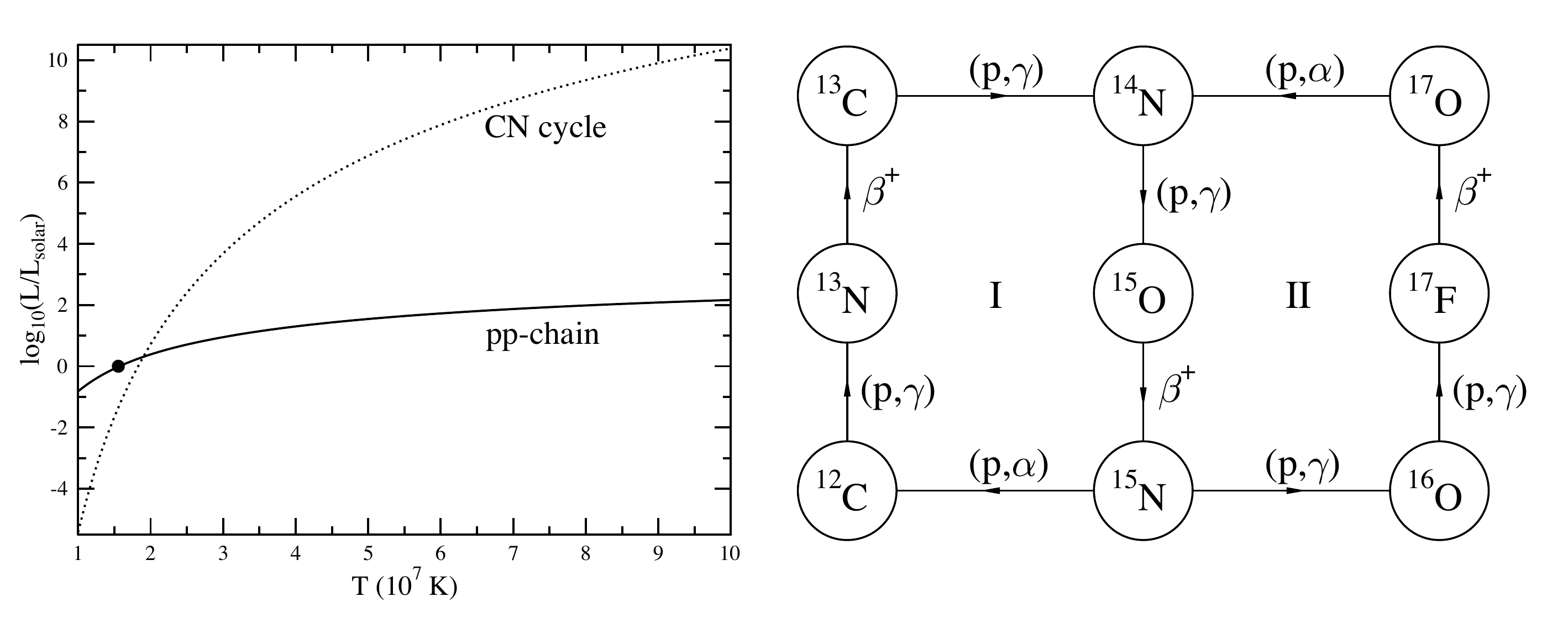}
\end{center}
\caption{The right panel shows the CNO bi-cycle for hydrogen burning.
The  left  compares  the energy  produced  in the  CN  cycle with  that
produced 
in the pp-chain, as a function of temperature T$_7$, measured
in units of 10$^7$ K.  The results are normalized to the 
pp-chain energy production at the  Sun's center and to solar metalicity,
assuming  equilibrium burning.  The  sharper  CN-cycle dependence  on
temperature 
is apparent.  If approximated as a  power law T$^x$, $x$ ranges between $\sim$
19 
and  $\sim$  22   over  the  range  of  temperatures   typical  of  the  Sun's
hydrogen-burning core. 
The dot marks the point corresponding to the Sun's center,
T$_7 \sim$ 1.57.}
\label{fig:two}
\end{figure}

But the  sharper temperature dependence of a second mechanism,  the CNO bi-cycle, 
is needed to account for
the structure  of more massive main-sequence  stars.  Unlike the pp-chain,  the CNO
bi-cycle  (Fig.~\ref{fig:two})  is a  secondary process:  the catalysts  for H
burning  are the  pre-existing metals.   Thus the  CNO contribution  to energy
generation is  proportional to stellar metalicity.   The CN-cycle, denoted by I  in Fig.~\ref{fig:two},
operates in the Sun.  The cycle conserves  the number abundance,
but alters the distribution of metals as it burns into equilibrium, eventually
achieving equilibrium abundances proportional to the inverses of the respective
rates.

While responsible for only $\sim$ 1\% of solar energy generation, the CN cycle
produces measurable neutrino fluxes.  
The slowest, rate-controlling reactions in the CN cycle are $^{14}$N(p,$\gamma$)
and  $^{12}$C(p,$\gamma$).  The latter has reached equilibrium over most of
the solar core.  In fact, the rapid burning of $^{12}$C, as the cycle
seeks equilibrium at the start of the main sequence, likely drives convective
mixing of the early solar core for $\sim 10^8$ y.
In contrast, the $^{14}$N  lifetime is shorter than the age of  the Sun only in regions where $T_7
\gsim 1.33$, which corresponds to $R \lsim  0.1 R_\odot$ in today's Sun,
or equivalently the central 7\% of the  Sun by mass.
Consequently, over a significant portion  of the outer core, $^{12}$C has been
converted   to  $^{14}$N,  but   further  reactions   are  inhibited   by  the
$^{14}$N(p,$\gamma$) bottleneck.

The BSP08(GS) SSM \cite{bps08} -- which employs recent updates of
metal abundances and of the $^{14}$N(p,$\gamma$) S-factor  -- finds a
modest  CN-cycle  contribution  to   solar  energy  generation  of  0.8\%  and
corresponding neutrino fluxes
\begin{displaymath}
^{13}\mathrm{N} (\beta^+)^{13}\mathrm{C}~~E_\nu \lsim 1.199 
\mathrm{~MeV}~~\phi = (2.93^{+0.91}_{-0.82}) \times 10^8/\mathrm{cm^2s}
\end{displaymath} 
\begin{displaymath}
^{15}\mathrm{O} (\beta^+)^{15}\mathrm{N} ~~E_\nu \lsim 1.732
\mathrm{~MeV}~~\phi          =          (2.20^{+0.73}_{-0.63})          \times
10^8/\mathrm{cm^2s}. 
\end{displaymath}
The ranges reflect conservative abundance  uncertainties  as defined
empirically in \cite{bs05}.
The first  reaction is part of the  path from $^{12}$C to  $^{14}$N, while the
latter  follows $^{14}$N(p,$\gamma$).   Thus neutrinos  from  $^{15}$O $\beta$
decay  are produced  in the  central core:  95\% of  the flux  comes  from the
CN-equilibrium region, described above.   About 30\% of the $^{13}$N neutrinos
come from outside  this region, primarily because of  the continued burning of
primordial  $^{12}$C: this  accounts for  the  somewhat higher  flux of  these
neutrinos. 

The  SSM makes  several  reasonable assumptions,  including local  hydrostatic
equilibrium (the balancing of the gravitational force against the gas pressure
gradient), energy  generation by proton  burning, a homogeneous  zero-age Sun,
and boundary conditions  imposed by the known mass,  radius, and luminosity of
the  present Sun.   It assumes  no significant  mass loss  or  accretion.  The
homogeneity assumption  allows the primordial  core metalicity to be  fixed to
today's surface  abundances.  Corrections for  the effects of diffusion  of He
and the  heavy elements  over 4.57  b.y. of solar  evolution are  included, and
generally been helpful in improving  the agreement between SSM predictions and
parameters probed in helioseismology.

The SSM postulate of a homogeneous zero-age  Sun is based on  the assumptions
1) that the early pre-main-sequence Sun passed through a fully convective, highly luminous
Hayashi  phase,  mixing  the  Sun; and 2) that no chemical differentiation in the
gas accreted onto the Sun occurred after this phase, when the Sun develops a
radiative core distinct from the convective surface zone.

Solar surface  abundances are known, determined from  analyses of photospheric
atomic  and  molecular spectral  lines.   Traditionally  the associated
atmospheric  modeling has  been done  in one  dimension, in  a time-independent
hydrostatic  analysis that incorporates  convection via  mixing-length theory.
But  much improved  3D  models of  the  solar atmosphere  have been  developed
recently  to treat  the radiation-hydrodynamics  and time  dependence  of this
problem.   This  3D analysis led to a revision in
solar  metalicity from the  previous standard, Z=0.0169  \cite{gs98}, to
Z=0.0122  \cite{ags05},  thus  altering  SSM  predictions. Hereafter  we
denote these as the GS and and AGS abundances, respectively. 

The predictions of solar models that use  the GS solar composition, the most up  to date of which
is the BPS08(GS) \cite{bps08} but including also the BP00 \cite{bpb00}, BP04
\cite{bp04}  and BS05(OP)  \cite{bsb05} models,  are in  excellent agreement
with helioseismology.   But those computed with the revised
abundances are in 
much poorer  agreement, with  discrepancies exceeding 1\%  in the  region just
below  the   convective  zone  (R  $\sim  0.65-0.70   {\rm  R}_\odot$). 
Associated  properties of the SSM,  such as
the depth of the convective zone and the surface He abundance, are also now in
conflict  with helioseismology.   As discussed  in \cite{bsb_mc}, the
discrepancies   are   significantly   above   measurement  and   solar   model
uncertainties. 

The   reduced  core   opacity  also   lowers   the  SSM   prediction  of   the
temperature-dependent $^8$B  neutrino flux by about 20\%:  the predicted $^8$B
flux using  the GS abundances  and Opacity Project  \cite{opacity} opacities
(model BPS08(GS)) is 5.95 $\times$ 10$^6$/cm$^2$s, 
 which drops  to 4.72  $\times$ 10$^6$/cm$^2$s when  AGS abundances  are used
(model BPS08(AGS)).  These results can be compared to the $^8$B neutrino flux
deduced from  the NCD-phase  SNO data set  of [5.54  $^{+0.33}_{-0.31}$ (stat)
$^{+0.36}_{-0.34}$ (sys)]    $\times$     10$^6$/cm$^2$s    \cite{sno_ncd}.

\section{The Sun as a Calibrated Laboratory}

It has been  recognized for many years  that a measurement of  the CN-cycle solar
neutrino flux would, in principle, determine the metalicity of this core zone,
allowing  a comparison with  abundance determined  from the  solar atmosphere.
In the past several years  new developments have occurred that may
make such a measurement practical:
\begin{itemize}

\item  Accurate  calibrations  of  the  solar  core  temperature  by  SNO  and
SK;

\item Tight constraints on the  oscillation parameters and matter effects that
determine the flavor content of the CN and $^8$B neutrino fluxes;

\item  Recent  measurements of  the  controlling  reaction  of the  CN  cycle,
$^{14}$N(p,$\gamma$),  that  have significantly  reduced  the nuclear  physics
uncertainties affecting SSM predictions of CN-cycle fluxes; and

\item New ideas for high-counting  rate experiments that would be sensitive to
CN-cycle  neutrinos,  and from  which  reliable  terrestrial  fluxes could  be
extracted.

\end{itemize}
The  analysis of \cite{whas}, which examined whether CN neutrino experiments could
place a significant constraint on the solar core's primordial metalicity,
used  previous  SSM  work  in  which  the  logarithmic  partial
derivatives $\alpha(i,j)$  for each neutrino  flux $\phi_i$ are  evaluated for
the SSM input parameters $\beta_j$,
\begin{equation}
\alpha(i,j)  \equiv  {\partial   \ln{\left[  \phi_i/\phi_i(0)  \right]}  \over
\partial \ln{\left[ \beta_j / \beta_j(0)\right]}},
\end{equation}
where  $\phi_i(0)$  and  $\beta_j(0)$   denote  the  SSM  best  values.   This
information, in combination with  the assigned uncertainties in the 19 $\beta_j$ of the SSM,
then  provides  an  estimate of  the  uncertainty  in  the SSM  prediction  of
$\phi_i$.   In particular,  crucial to  the current  analysis is the dependence
\cite{bs05} on
the  mass  fractions  (measured  relative  to  hydrogen)  of  different  heavy
elements,
\begin{equation}
\beta_j =  \mathrm{mass~fraction~of~element~j \over mass~fraction~of~hydrogen}
\equiv X_j.
\end{equation}
Having this information  not as a function of the  overall metalicity $Z$, but
as  a  function  of the  individual  abundances,  allows  one to  separate  the
``environmental'' effects  of the  metals in the  solar core from  the special
role of  primordial C and N as  catalysts for the CN  cycle.  By environmental
effects  we mean  the influence  of the  metals on  the opacity  and  thus the
ambient  core  temperature, which  controls  the  rates of  neutrino-producing
reactions of both the pp-chain and CN cycle.  In \cite{whas}
the  temperature-dependent  $^8$B  neutrino  flux  was used to calibrate the
environmental  effects  of  the  metals  and of  other  SSM  parameters,  thus
isolating the  special CN-cycle  dependence on primordial  C+N.  This
primordial abundance can  be expressed, with very little  residual solar model
uncertainty, in  terms of the measured  $^8$B neutrino flux  and nuclear cross
sections that have been determined in  the laboratory.

The  partial derivatives  allow one  to define  the power-law  dependencies of
neutrino fluxes, relative to the SSM best-value prediction $\phi_i(0)$
\begin{equation}
\phi_i   =   \phi_i(0)  \prod_{j=1}^N   \left[   {\beta_j  \over   \beta_j(0)}
\right]^{\alpha(i,j)}
\label{eq:prod}
\end{equation}
where the product extends over  $N=19$ SSM input parameters.  This expression can
be  used  to  evaluate  how  SSM  flux  predictions  will  vary,  relative  to
$\phi_i(0)$, as the  $\beta_j$ are varied.  Alternatively, the  process can be
inverted:  a flux  measurement  could in  principle  be used  to constrain  an
uncertain input parameter.

The baseline SSM calculation used in \cite{whas}, BPS08(AGS) \cite{bps08},
employed the recently determined AGS
abundances for  the volatile  elements C, N,  O, Ne,  and Ar, rather  than the
previous  GS standard  composition. It  should be  noted that  AGS  includes a
downward revision by 0.05 dex of  the Si photospheric abundance compared to GS
and,  accordingly, a  similar  reduction in  the  meteoritic abundances.   The
partial derivatives needed in the present calculation are summarized in
Tables \ref{table:one} (solar model parameters and nuclear cross sections) and
\ref{table:two} (abundances).

The SSM  estimate  of  uncertainties  in the  various  solar
neutrino fluxes 
$\phi_i$  can  be  obtained  by  folding  the  partial  derivatives  with  the
uncertainties in the underlying $\beta_j$.  In particular, it is convenient to
decompose Eq.~( \ref{eq:prod}) into its  dependence on solar  parameters and non-CN
metals, nuclear S-factors, and the primordial C and N abundances,
\begin{equation}
\phi_i =  \phi_i^{SSM}  \prod_{j \in \mathrm{\{Solar~and~Metals  \neq  C,N\}}} \left[  {\beta_j \over \beta_j(0)}
\right]^{\alpha(i,j)} 
\prod_{j   \in  \mathrm{\{Nuclear\}}}   \left[   {\beta_j  \over   \beta_j(0)}
\right]^{\alpha(i,j)}  \prod_{j \in  \mathrm{\{C,N\}}}  \left[ {\beta_j  \over
\beta_j(0)} \right]^{\alpha(i,j)}
\label{eq:prod2}
\end{equation}
where the  first  term will  be  designated the  ``environmental''
uncertainty --  SSM solar parameters and metal abundances  that primarily influence
neutrino flux predictions through changes they induce in the core temperature.
These are, respectively, the uncertainties in the photon luminosity $L_\odot$,
the  mean  radiative opacity,  the  solar age,  and  calculated  He and  metal
duffusion; and the fractional abundances of O, Ne, Mg, Si, S, Ar, and Fe.  The
estimated  1$\sigma$ fractional  uncertainties for  the these parameters
are given in \cite{whas}.
The abundances of Mg,
Si, S,  and Fe are meteoritic, while those of the volatile elements  C,   N,  O,   Ne,   and Ar
are photospheric.   

The next term  contains the effects of nuclear  cross section uncertainties on
flux predictions.  The $\beta_j$ are  the S-factors for p+p ($S_{11}$), $^3$He
+  $^3$He ($S_{33}$),  $^3$He+$^4$He ($S_{34}$),  p +  $^7$Be ($S_{17}$),  e +
$^7$Be ($S_{e7}$),  and p +  $^{14}$N ($S_{114}$).  The  estimated 1$\sigma$
fractional  uncertainties are also given in \cite{whas}.

The last  term is the contribution of  the primordial C and  N abundances.  As
Table   \ref{table:two}  shows,  pp-chain   neutrino  fluxes   are  relatively
insensitive to variations  in these abundances, as the  heavier nuclei like Fe
have a more important influence on the core opacity.  But the expected, nearly
linear  response  of  the  $^{13}$N  and $^{15}$O  neutrino  fluxes  to  these
abundances is apparent.   These are the abundances we  would like to constrain
by a  future measurement of the  $^{13}$N and $^{15}$O  solar neutrino fluxes.
Such  a measurement  begins to  be of  interest if  these abundances  could be
determined with an accuracy significantly better than 30\%.  

Were  one  to vary  the  11  SSM  parameters designated  as  ``environmental''
according to their assigned uncertainties (taking them to be uncorrelated), a
7.5\% SSM  net uncertainty in $\phi(^{13}\mathrm{N})$ would  be obtained.  
This uncertainty would be one of dominant ones in the analysis of present
interest.  But, as discussed in \cite{whas} in more detail, these environmental
uncertainties influence neutrino fluxes through their impact on the core
temperature, regardless of the details of which parameters are
varied.  Consequently, there are strong environmental correlations between different
neutrino fluxes, allowing one to form ratios of fluxes that are much less sensitive
to environmental effects.  Furthermore these correlations remain valid for parameter
variations far outside accepted SSM uncertainties.

\begin{figure}
\begin{center}
\includegraphics[width=16cm]{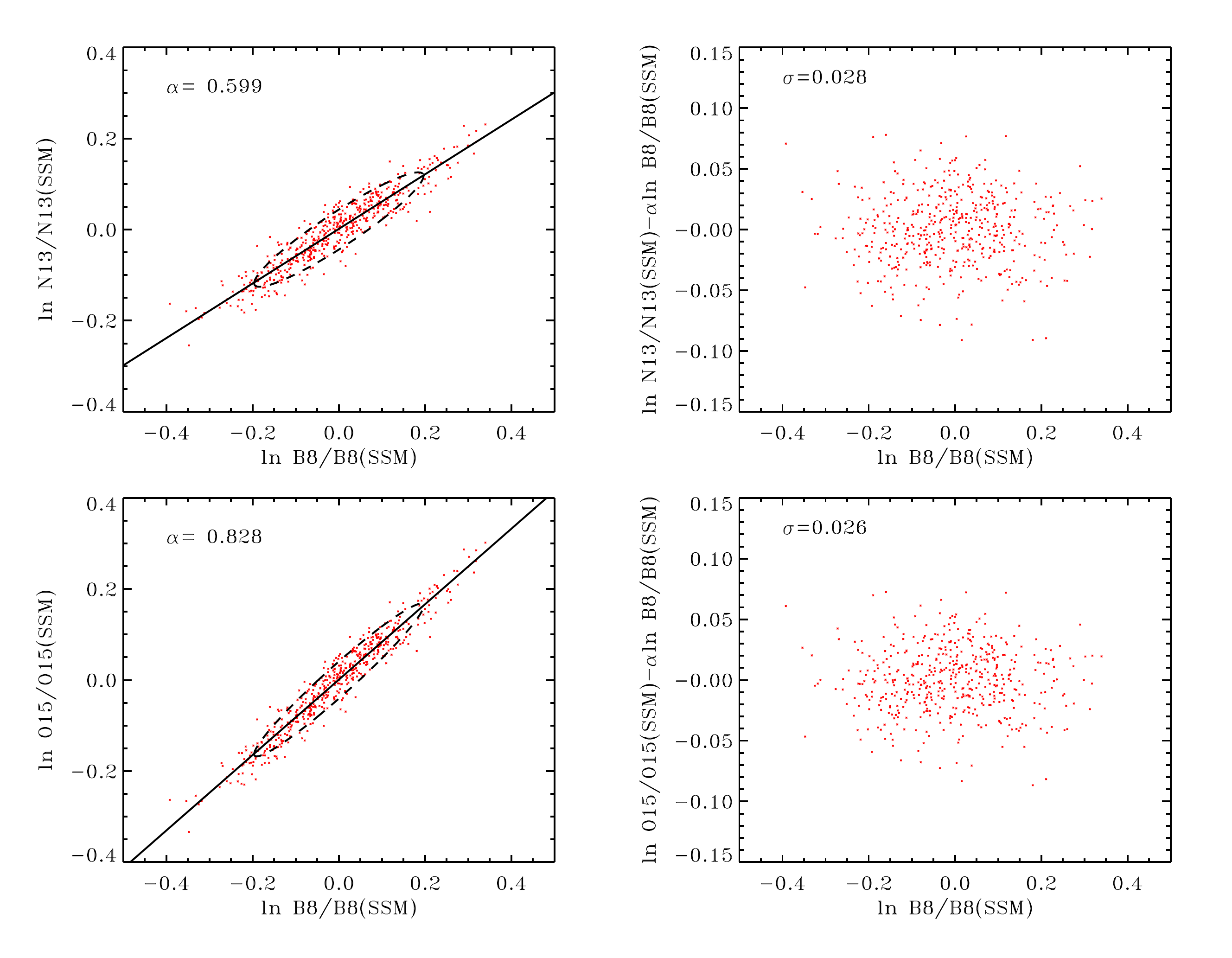}
\caption{Results  from  SSM Monte  Carlo  simulations  in which the  11
  environmental parameters  (see text) have  been varied. The two  left panels
  show  the correlations between  the $^8{\rm  B}$ flux  and the  two CN-cycle
  neutrino  fluxes  $^{13}{\rm N}$  and  $^{15}{\rm  O}$.  The slopes  of  the
  correlations are  given in  the plots, together  with the  68.3\% confidence
  level contours.  The panels on the right show  the residuals  from the
  fits,  2.8\% and  2.6\% for  the $^{13}{\rm  N}$ and  $^{15}{\rm  O}$ fluxes
  respectively $-$ the residual environmental  uncertainty that remains after
  making use of the $^8$B flux constraint.}
\end{center}
\label{fig:three}
\end{figure}

This is illustrated in the SSM Monte Carlo tests represented in Fig.~\ref{fig:three}.
One finds that a flux ratio such as
\begin{equation}
{\phi(^{15}\mathrm{O}) \over \phi(^8\mathrm{B})^K},
\end{equation}
where the exponent $K \sim 0.828$ is taken from the fit shown in
Fig.~\ref{fig:three}, is much less sensitive to environmental
uncertainties than either the numerator or denominator separately.  This
is apparent from the entries in Tables \ref{table:one} and \ref{table:two}.
As $\phi(^8\mathrm{B})$ can be taken from SK
and SNO measurements, most of the environmental uncertainty
in predicting $\phi(^{15}\mathrm{O})$ can be eliminated.
Effectively, $\phi(^8\mathrm{B})$ becomes a ``thermometer" constraining core
temperature changes induced by varying the environmental $\beta_j$.
In this way one obtains a more precise relationship between the CN flux, 
a quantity that should be measured
quite accurately in next-generation experiments like SNO+, and the core abundances
of C and N.

\begin{table}
\caption{Partial  derivatives  $\alpha(i,j)$  of  neutrino  fluxes  with
    respect  to  solar environmental  and  nuclear  cross section  parameters.
    \label{table:one}} 
\begin{tabular}{|l|cccc|cccccc|}
\hline
 & \multicolumn{4}{|c|}{Environmental $\beta_j$} &
\multicolumn{6}{c|}{Nuclear $\beta_j$} \\
Source  & $L_\odot$  &  Opacity &  Age  & Diff.~&  $S_{11}$  & $S_{33}$  &
$S_{34}$ & $S_{17}$ & $S_{e7}$ & $S_{114}$ \\ 
\hline
$\phi$($^8$B) & 7.16 & 2.70 & 1.38 & 0.28 & -2.73 & -0.43 & 0.85 & 1.0 & -1.0 &
-0.020 \\ 
$\phi$($^{13}$N) & 4.40 & 1.43 & 0.86 & 0.34 & -2.09 & 0.025  & -0.053 & 0.0 &
0.0 & 0.71 \\ 
$\phi$($^{13}$N)/$\phi$($^8$B)$^{0.599}$ & 0.11 & -0.19  & 0.03 & 0.17 & -0.45
& 0.28 & -0.56 & -0.60 & 0.60 & 0.72 \\ 
$\phi$($^{15}$O) & 6.00 & 2.06 & 1.34 & 0.39 & -2.95 & 0.018 & -0.041 & 0.0 &
0.0 & 1.00 \\ 
$\phi$($^{15}$O)/$\phi$($^8$B)$^{0.828}$ & 0.07 & -0.18  & 0.20 & 0.16 & -0.69
& 0.37 & -0.74 & -0.83 & 0.83 & 1.02 \\ 
\hline
\end{tabular}
\end{table}

\begin{table}
\caption{Partial  derivatives  $\alpha(i,j)$  of  neutrino  fluxes  with
    respect    to   fractional    abundances   of    the    primordial   heavy
    elements. \label{table:two}} 
\begin{tabular}{|l|cc|ccccccc|}
\hline
&      \multicolumn{2}{|c|}{C,       N       $\beta_j$}      &
  \multicolumn{7}{c|}{Environment Abundance $\beta_j$} \\
Source & C& N & O & Ne & Mg & Si & S & Ar & Fe \\
\hline
$\phi$($^8$B) & 0.027 & 0.001 & 0.107 & 0.071 & 0.112 & 0.210 & 0.145 & 0.017
& 0.520 \\ 
$\phi$($^{13}$N) &  0.874 & 0.142 &  0.044 & 0.030 &  0.054 & 0.110  & 0.080 &
0.010 & 0.268 \\ 
$\phi$($^{13}$N)/$\phi$($^8$B)$^{0.599}$ &  0.858 & 0.141 & -0.020  & -0.013 &
-0.013 & -0.016 & -0.007 & 0.000 & -0.043 \\ 
$\phi$($^{15}$O) &  0.827 & 0.200 &  0.071 & 0.047 &  0.080 & 0.158  & 0.113 &
0.013 & 0.393 \\ 
$\phi$($^{15}$O)/$\phi$($^8$B)$^{0.828}$ &  0.805 & 0.199 & -0.018  & -0.012 &
-0.013 & -0.016 & -0.007 & -0.001 & -0.038 \\ 
\hline
\end{tabular}
\end{table}

\section{The Analysis for Elastic Scattering and Neutrino Oscillations}

The analysis requires a number of steps that will be summarized here,
as more detail can be found in Ref. \cite{whas}.
The total $^8$B flux (the instantaneous solar flux), normalized to the SSM
best value, can be related to rates that would be measured in terrestrial detectors by
\begin{equation}
{\phi(^8\mathrm{B})   \over  \phi^{SSM}(^8\mathrm{B})} =  {\phi(^8\mathrm{B})
\langle  \sigma^{SK}(^8\mathrm{B}, \delta m_{12}^2,\theta_{12})  \rangle \over
\phi^{SSM}(^8\mathrm{B})      \langle     \sigma^{SK}(^8\mathrm{B},     \delta
m_{12}^2,\theta_{12})      \rangle }    \equiv
{R^{SK}_{exp}(^8\mathrm{B})     \over     R^{SK}_{cal}(^8\mathrm{B},    \delta
m_{12}^2,\theta_{12})}
\label{eq:fluxratio}
\end{equation}
Here $\langle  \sigma^{SK} \rangle$ is  an effective cross section  that takes
into  account  all of  the  neutrino  flavor  and detector  response issues  (trigger
efficiencies,  resolution,  cross  section  uncertainties, oscillations, etc.) that
determine  the   relationship  between  a  measured  detector   rate  and  the
instantaneous  solar flux.   The numerator  of  the ratio  on the  right is  a
directly   measured  experimental   quantity:  the   SK  elastic
scattering rate for producing  recoil electrons with apparent energies between
5.0  and  20 MeV,  per  target  electron per  second.   The  denominator is  a
theoretical quantity,  computed by folding the SSM best-value $^8$B flux produced
in the Sun with the  cross  section  for  ($\nu_x,e)$  elastic  scattering,
averaged  over   a  normalized  $^8$B  spectrum,  defined   for  the  specific
experimental  conditions of  SK,  and including  the effects  of
flavor mixing that alter the flux during transit from the Sun to the detector.   
This cross section is calculated from  laboratory measurements of detector properties, the
$\beta$ decay  spectrum, the underlying neutrino-electron  cross sections, and
most  critically, the  parameters governing  oscillations.  We  describe these
factors below.

The  experimental   rate  comes  from   the  1496  days  of   measurements  of
SK I \cite{hosaka}.  From the SK I rate/kiloton/year
\begin{equation}
520.1        \pm         5.3        \mathrm{(stat)}        ~{}^{+18.2}_{-16.6}
    \mathrm{(sys)}~\mathrm{kton^{-1}~y^{-1}}.
\label{eq:rate1}
\end{equation}
we find $R^{SK}_{exp}(^8\mathrm{B})$,
\begin{equation}
4.935    \pm    0.05   \mathrm{(stat)}~{}^{+0.17}_{-0.16}\mathrm{(sys)}~\times
  10^{-38}~\mathrm{{electron}^{-1} s^{-1}}
\end{equation}
(or  $\sim$ 0.049  Solar Neutrino  Units, or  SNUs).  The  dominant systematic
error  includes  estimates  for  the  energy  scale  and  resolution,  trigger
efficiency, reduction, spallation dead time,  the gamma ray cut, vertex shift,
background  shape  for  signal  reduction, angular  resolution,  and  lifetime
uncertainties.  The  combined statistical and  systematic error is  $\sim \pm$
3.6\%.

To evaluate the denominator in  Eq.~(\ref{eq:fluxratio}) we need the suitably
averaged cross section, defined for the window used by the SK I collaboration,
\begin{eqnarray}
 \langle \sigma^{SK}(^8\mathrm{B},  \delta m_{12}^2,\theta_{12}) \rangle= \int
d  E_\nu  \phi_{norm}^{^8\mathrm{B}}(E_\nu) 
\left[       P_{\nu_e}(E_\nu,\delta       m^2_{12},\sin^2{2      \theta_{12}})
\int_{T=0}^{T^{max}(E_\nu)}   dT    ~\sigma_{\nu_e}^{es}(T)   \right.   ~~~~&&
\nonumber    \\   \left.    +    ~P_{\nu_\mu}(E_\nu,\delta   m^2_{12},\sin^2{2
\theta_{12}}) \int_{T=0}^{T^{max}(E_\nu)} dT ~\sigma_{\nu_\mu}^{es}(T) \right]
 \int_{5~\mathrm{MeV}}^{20~\mathrm{MeV}}  d
\epsilon_a                                     f_{\mathrm{trig}}(\epsilon_a)
\rho(\epsilon_a,\epsilon_t=T+m_e)
\label{eq:theorypart}
\end{eqnarray}
where  $\phi_{norm}^{^8\mathrm{B}}(E_\nu)$ is  the  normalized $^8$B  neutrino
spectrum.   Equation  (\ref{eq:theorypart})  involves  an  integral  over  the
product    of   this   spectrum    and   the    energy-dependent   oscillation
probabilities.  ($P_{\nu_e}+P_{\nu_\mu}$=1, assuming oscillations  into active
flavors.  $P_{\nu_\mu}$ can be defined as the oscillation probability to heavy
flavors, if  the effects  of three flavors  are considered.)  A  given $E_\nu$
fixes the range of kinetic energies  $T$ of the scattered electron, over which
an integration  is done; in the  laboratory frame $T^{max}  = 2 E_\nu^2/(m_e+2
E_\nu)$.   The  integrand  includes  the  elastic  scattering  cross  sections
$\sigma^{es}(T)$   for   electron   and   heavy-flavor   neutrinos   and   the
SK  resolution   function  $\rho(\epsilon_a,\epsilon_t)$,  where
$\epsilon_t=T+m_e$  is the true  total electron  energy while  $\epsilon_a$ is
apparent energy, as deduced from the number of phototube hits in the detector.
Finally,   an  integral   must  be   done  over   the  window   used   by  the
experimentalists, apparent electron energies $\epsilon_a$ between 5 and 20
MeV.  The  deduced counting  rate includes the  triggering probability  that a
event of apparent energy $\epsilon_a$ will be recorded in the detector. 
Resolution and triggering functions for SK are given in \cite{whas}. 
        
Similarly, the CN-cycle  neutrino response for a detector like SNO+ is
\begin{equation}
{\phi(^{15}\mathrm{O})       \over       \phi^{SSM}(^{15}\mathrm{O})}  \equiv     {      ~R^{B/S}_{exp}(\mathrm{^{15}\mathrm{O}})     \over
R^{B/S}_{cal}(^{15}\mathrm{O}, \delta  m_{12}^2,\theta_{12})} =
{R_{exp}^{B/S}(\mathrm{CN})/(1             +\alpha(0.8,1.3))             \over
R^{B/S}_{cal}(^{15}\mathrm{O}, \delta m_{12}^2,\theta_{12})) }
\label{eq:fluxratio2}
\end{equation}
where the experimental rate for $^{15}$O neutrinos has been written in terms of
the  total  CN-neutrino  rate  $R^{B/S}_{exp}(\mathrm{CN})$ by  introducing  a
correction   factor    $\alpha \sim 0.12$ that accounts for the $^{13}$N neutrino
contribution. As discussed in \cite{whas}, $\alpha$ can be measured
in principle, but can also be evaluated from theory, with negligible uncertainty.
No  measurement   of
$R^{B/S}_{exp}(\mathrm{CN})$   currently   exists,   of  course.    But   such
measurements could be made in  Borexino or SNO+, existing or planned detectors
using large  volumes  of  organic  scintillator and placed quite  deep
underground.  A
window  for the  apparent  kinetic energy  $T$  of the  scattered electron  of
0.8-1.3 MeV has been discussed by the Borexino group.  As the $^7$Be 0.866 MeV
line  corresponds  to  $T^{max}  \sim$  0.668 MeV,  this  window  would  limit
contamination from $^7$Be neutrino recoil electrons.  

As discussed in \cite{whas}, one can use Eqs.~(\ref{eq:fluxratio})~and~(\ref{eq:fluxratio2})
in expressions based on Eq.~(\ref{eq:prod2}), then on dividing obtain
\begin{eqnarray}
 {R_{exp}^{B/S}(\mathrm{CN})   \over   R^{B/S}_{cal}(^{15}\mathrm{O},   \delta
m_{12}^2,\theta_{12}))  }  = 
(1.120     \pm     0.003)     \left[     {R^{SK}_{exp}(^8\mathrm{B})     \over
R^{SK}_{cal}(^8\mathrm{B},            \delta            m_{12}^2,\theta_{12})}
\right]^{0.828}~~~~~~~~~~ ~~~~~~~~~~ \nonumber   \\  \times   \left[   1   \pm   2.6\%
(\mathrm{resid.~envir.})  \pm 7.6\%  (\mathrm{nuclear})  \right] 
 \left(     {X(^{12}\mathrm{C})    \over    X(^{12}\mathrm{C})_{SSM}}
\right)^{0.805}  \left(  {X(^{14}\mathrm{N})  \over  X(^{14}\mathrm{N})_{SSM}}
\right)^{0.199}.
\label{eq:final}
\end{eqnarray}
Effectively the SK rate has been used to limit SSM
``environmental'' uncertainties,  leaving an error budget
dominated  by  the  nuclear   physics.   But this source of error is under laboratory
control and will  be reduced as nuclear
reaction measurements continue.   The last two terms are the primordial
abundances one would  like to constrain.  The role of the  SSM in this equation
is to  define a set  of parameters  and thus a  set of reference  rates, about
which  we then  explore possible  variations.  Those  variations  generate the
environmental   and   nuclear   uncertainties   discussed  above.

The   $R_{cal}$   factors   in   Eq.~(\ref{eq:final})   contain   additional
uncertainties discussed in \cite{whas}, including one important one, that associated
with neutrino  oscillations. Apart  from the dependence  on the  solar density
profile, this should be considered a laboratory uncertainty, analogous to nuclear
cross sections.  Uncertainties in oscillation parameters will continue to be refined
by astrophysical, accelerator and reactor measurements. 

The  LMA parameter  uncertainties  in SK  and Borexino/SNO+  are
anti-correlated.  Most of the  low-energy $^{15}$N neutrinos do not experience
a level  crossing, residing instead  in a portion  of the MSW plane  where the
oscillations are close to the vacuum oscillation limit,
\begin{equation}
P_{\nu_e}(E_\nu) \rightarrow 1 - {1 \over 2} \sin{2 \theta_{12}},
\end{equation}
so that an increase  in  the  vacuum mixing  angle  $\theta_{12}$ decreases  the
$\nu_e$ survival  probability.  The higher energy $^8$B  neutrinos are largely
within  the MSW  triangle,  described  by an  adiabatic  level crossing.   The
limiting behavior for an adiabatic crossing is
\begin{equation}
P_{\nu_e}(E_\nu) \rightarrow {1 \over 2} (1 -\cos{2 \theta_{12}})
\end{equation}
so that an increase in $\theta_{12}$ increases the survival probability.  This
anti-correlation thus leads to larger effects in the ratio.

The impact of this uncertainty on Eq.~(\ref{eq:final}) was analyzed in \cite{whas}, using
the  allowed  regions for  $\theta_{12}$  and  $\delta  m_{12}^2$ obtained  in
KamLAND's  combined analysis, yielding
\begin{equation}
{R^{B/S}_{cal}(^{15}\mathrm{O},     \delta     m_{12}^2,\theta_{12}))    \over
 R^{SK}_{cal}(^8\mathrm{B},     \delta     m_{12}^2,\theta_{12})^{0.825}}    =
 (1 \pm 0.049)\left[ {R^{B/S}_{cal}(^{15}\mathrm{O},
 \delta   m_{12}^2,\theta_{12}))   \over   R^{SK}_{cal}(^8\mathrm{B},   \delta
 m_{12}^2,\theta_{12})^{0.825}} \right]^{BV}
\end{equation}
where $BV$ denotes the SSM best-value ratio.

Thus  the overall  uncertainty budget  in Eq.~(\ref{eq:final}) includes the
experimental uncertainty of the SK  ``thermometer'' of 3\%,  residual solar
environmental uncertainties at 2.6\%, LMA parameter uncertainties at 4.9\%,
and nuclear S-factor uncertainties of 7.6\%.
The overall  uncertainty in the ``theoretical'' relationship
between a future SNO+ or Borexino CN-neutrino flux and core C/N metals is thus
about 9.6\%.  As  the nuclear physics uncertainty dominates, one
would expect this relationship to become more precise when ongoing analyses of
data obtained by the LUNA collaboration and others for
$^{14}$N(p,$\gamma$) are completed.  An appropriate goal
would  be 3.5\%  in this  S-factor, a  30\% improvement.   The  uncertainty in
$^{14}$N(p,$\gamma$)  would  no  longer  dominate the  nuclear  physics  error
budget, but instead would be comparable to the contributions from S$_{33}$ and
S$_{34}$.  However, the current 9.6\% uncertainty is not a bad starting point,
as first-generation CN-cycle neutrino experiments are expected to measure this
flux to an accuracy of about  10\%.  That is, the theoretical uncertainty will
not dominate the experimental uncertainty, even without anticipated nuclear
physics improvements.

\section{Future Experiments and Summary}

The work reported here was motivated in part by new detectors
that  might  enable  a  high-statistics  measurement  of  the  CN-cycle
neutrinos.    Two possibilities are Borexino and SNO+, detectors based on ultra-clean   organic
scintillation liquids.  Borexino, which operates within Gran Sasso,
must deal with a serious background, 
cosmogenic $^{11}$C (a $\beta^+$  source).  The collaboration has discussed
a possible triple-coincidence veto \cite{franco} to limit this background.

\begin{figure}
\begin{center}
\includegraphics[width=13cm]{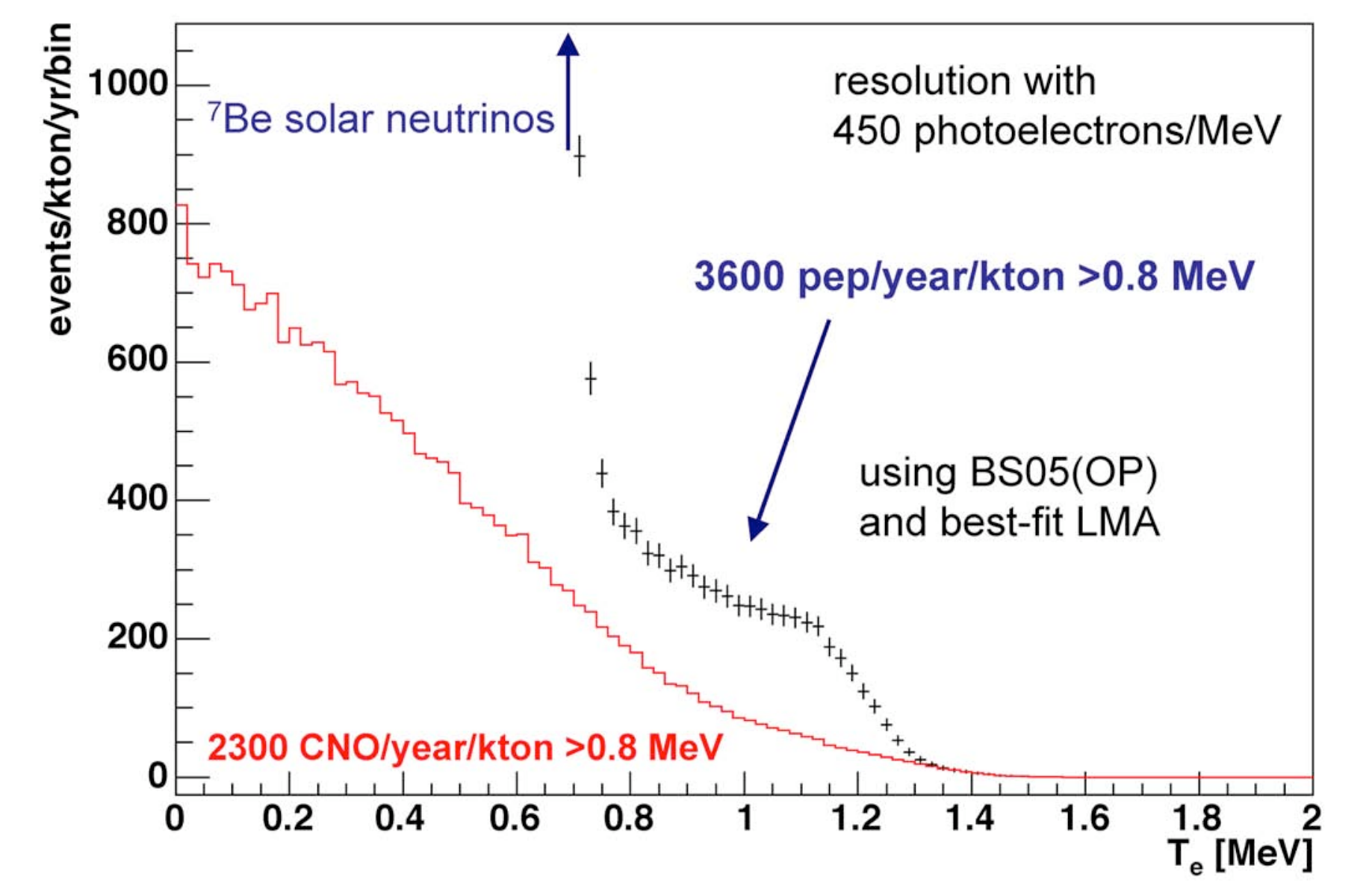}
\end{center}
\caption{A simulation of the
 $^7$Be, pep, and CNO electron recoil spectrum expected in SNO+.
This figure is due to M. C. Chen \cite{chen}.}
\label{fig:CNO_SNOPlus}
\end{figure}

Because of SNOLab's 6.0 km.w.e. depth, $^{11}$C will be much less 
troublesome in SNO+, an experiment that will use the SNO cavity and about
three times more scintillator than Borexino.
Figure \ref{fig:CNO_SNOPlus} shows a simulation of the expected SNO+ response,
performed by the experimenters (Chen, private communication).  (Note that
the simulation is based on  the  BS05(OP) SSM and the best-fit LMA solution 
to  the solar  neutrino problem, rather than the updated BPS08(AGS) used in this
paper.)  The CN-neutrino  event rate  for  an energy
window  above 0.8 MeV  was found  to be  2300 counts/year.   The experimenters
concluded that  SNO+ could  determine the CN-neutrino  rate to an  accuracy of
approximately 10\%, after three years of running \cite{chen}.  This is an
appropriate  goal  for  such  a  first-generation  CN-cycle  neutrino
measurement, as it  would approach the accuracy with which  that flux could be
related theoretically to  the Sun's primordial core C and  N abundances, as
argued in this paper.

One main point of this talk is that if a 10\% measurement can be made, an analysis
could be done that limits theoretical uncertainties to below this level, provided one
uses SK as a thermometer to eliminate environmental
uncertainties.  Indeed, both of the limiting errors in such an analysis
can be addressed in future laboratory measurements, namely
nuclear  cross sections ($\sim$ 7.6\%) and 
LMA oscillation  parameters ($\sim$  4.9\%).  One
goal should be  the reduction of these error bars to the level  of uncertainty of the
SK thermometer, $\sim$ 3\%.

In my view, the primary motivation for such a measurement is to test the SSM
assumption of a homogeneous zero-age-main-sequence Sun, given 
recent revisions of metal abundances derived from analyses of photospheric
absorption lines.   One could try to reconcile helioseismology with lower photospheric
metal abundances by violated this SSM assumption,
adopting a two-zone Sun with a low-Z convective zone.

It is intriguing that a possible mechanism for diluting the convective zone
exists.  The  solar system's
primary reservoir for  metals, the gaseous giant planets, formed 
late in the evolution of the solar nebula,
incorporating an excess of metal  estimated at 40-90 M$_\oplus$.  This mass is
similar to the deficit of metals in the convective zone, were one to interpret
the helioseismology/photospheric abundance discrepancy  in the most naive way.
The  raises a provocative question:  is it  possible that  the  process that
concentrated  metals in the  gaseous giants  also produced  a large  volume of
metal-depleted  gas that  subsequently was  accreted onto  the  Sun's surface?  If
so, late-stage accretion of depleted gas onto the Sun would have not only diluted
the convective zone, but generated a transition zone in the modern Sun's upper
radiative zone $-$ one that might alter helioseismology in that region.
One would also expect abundance anticorrelations between
the atmospheres of the Sun and gaseous giants.
While the  suggestion of  a common chemical  mechanism linking  the convection
zone  and  the gaseous  giants  is  speculative, it provides additional
motivation for  exploiting the CN neutrinos  as a quantitative  probe of solar
core metalicity.

\section{Acknowledgments}

This  work was  supported  in part  by  the  Office of  Nuclear
Physics,  US Department  of  Energy, under  grant  DE-FG02-00ER-41132.

\end{document}